\documentclass[preprint,onecolumn,nofootinbib]{revtex4}
\usepackage[colorlinks=true,linkcolor=blue,urlcolor=blue,filecolor=black,citecolor=red,pdfstartview=FitV,pdftitle={},pdfsubject={},pdfkeywords={},pdfpagemode=None,bookmarksopen=true]{hyperref}
\usepackage{graphicx}
\usepackage{amsmath}
\usepackage{amsfonts}
\usepackage{amssymb,ulem}
\usepackage{color,xcolor}%
\usepackage{CJK}
\usepackage{subfigure}
\usepackage{float}
\usepackage{appendix}

\usepackage{dcolumn}
\newcommand{\f}{\begin{equation}}
\newcommand{\ff}{\end{equation}}
\newcommand{\fa}{\begin{eqnarray}}
\newcommand{\ffa}{\end{eqnarray}}

\begin{document}
\title{Transport properties of a $3$-dimensional holographic effective theory with gauge-axion coupling}

\author{Yi-Lin Li}
\thanks{lyilin430@gmail.com}
\author{Xi-Jing Wang}
\thanks{xijingwang@yzu.edu.cn} 	
\author{Guoyang Fu}
\thanks{FuguoyangEDU@163.com}
\author{Jian-Pin Wu}
\thanks{jianpinwu@yzu.edu.cn}
\affiliation{Center for Gravitation and Cosmology, College of Physical Science and Technology, Yangzhou University, Yangzhou 225009, China
}

\begin{abstract}
	
In this paper, we implement a $3$-dimensional holographic effective theory with gauge-axion coupling. The analytical black hole solution is  worked out. We investigate the Direct current (DC) thermoelectric conductivities. 
A novel property is that DC electric conductivity for vanishing gauge-axion coupling is temperature dependent. It is different from that of $4$-dimensional axion model whose DC electric conductivity is temperature independent.
In addition, the gauge-axion coupling induces a metal insulator transition (MIT) at zero temperature.
The properties of other DC thermoelectric conductivities are also discussed. Moreover we find that the Wiedemann-Franz (WF) law is violated in our model.

\end{abstract}

\maketitle

\section{Introduction}
Since the electrons are collective excitations in $(1+1)$ dimensional condensed matter system, this system exhibits some peculiar properties, which are very different from that in higher dimensions \cite{Schulz:98}. In particular, $(1+1)$ dimensional system such as Luttinger liquid usually involves strong correlation, for which the perturbation theory fails. The bosonization approach is an alternative method to handle this problem \cite{Schulz:98}.

Recently, AdS/CFT (Anti-de Sitter/Conformal field theory) correspondence \cite{Maldacena:1997re,Gubser:1998bc,Witten:1998qj,Aharony:1999ti} provides a new way to attack the strongly correlated system and stand a chance of revealing the basic principle behind them. Some important progresses such as the holographic superconductor \cite{Hartnoll:2008vx}, holographic metal-insulator phase transition (MIT) \cite{Donos:2012js,Ling:2014saa,An:2020tkn} and (non-)Fermi liquid \cite{Liu:2009dm}, have been made by AdS$_4$/CFT$_3$ and AdS$_5$/CFT$_4$ correspondences. Various holographic dual systems from AdS$_3$/CFT$_2$ such as $(1+1)$ dimensional holographic superconductor models \cite{Ren:2010ha,Liu:2011fy,Li:2012zzb,Alkac:2016otd} and the properties of $(1+1)$ dimensional of holographic liquids \cite{Maity:2009zz,Faulkner:2012gt}, are also explored. However, the transport properties of $(1+1)$ dimensional holographic system with momentum dissipation are still absent in previous works.

The transports are the important characteristics of the strongly correlated systems. In holographic framework, the transport properties have also been widely explored, see the reviews \cite{Hartnoll:2009sz,Natsuume:2014sfa,Hartnoll:2016apf,Baggioli:2019rrs,Baggioli:2021xuv} and therein. After implementing the momentum relaxation, the holographic systems from AdS$_4$/CFT$_3$ and AdS$_5$/CFT$_4$ exhibit some exciting results such as the linear-T resistivity \cite{Hartnoll:2009ns,Davison:2013txa}, the quadratic-T inverse Hall angle \cite{Anderson,coleman1996should,Zhou:2015dha} and the MIT \cite{Mott:1968nwb,Ling:2016dck,Nishioka:2009zj}.

In this paper, we want to construct a $(1+1)$ dimensional holographic effective theory with gauge-axion coupling from AdS$_3$/CFT$_2$ and study its transport properties. We shall consider a simple implementation of momentum relaxation induced by an linearly spatial-dependent axionic field. The holographic axion model is first proposed from the higher dimensional AdS/CFT correspondence in \cite{Andrade:2013gsa}. This model breaks the translational symmetry by the linearly spatial-dependent axionic field, which induces the momentum relaxation in the dual boundary field theory, but retains the homogeneity of the background geometry. This way bypasses solving the partial differential equations differently from that in \cite{Horowitz:2012gs,Horowitz:2012ky,Horowitz:2013jaa}. In this simple model, we can obtain a finite DC conductivity thanks to the momentum relaxation.

Our paper is organized as follows. In section \ref{sec-EM-bh}, we construct a $3$-dimensional Einstein-Maxwell-axion model with gauge-axion coupling and then we work out the analytical black hole solution of this model. In section \ref{CC}, we calculate the DC thermoelectric transport coefficients of this holographic model and explore their properties. 
We present the conclusions and discussions in section \ref{conclu}.

\section{Holographic framework}\label{sec-EM-bh}

Following previous works of $4$-dimensional Einstein-Maxwell-axion model with gauge-axion coupling \cite{Gouteraux:2016wxj,Baggioli:2016pia,Li:2018vrz,Liu:2022bam,Zhong:2022mok}, we implement a $3$-dimensional holographic effective model with gauge-axion coupling, whose action is 
\begin{equation}
	S=\int d^{3} x \sqrt{-g}\left(R-2 \Lambda- X-\frac{1}{4} F^{2}-\frac{\mathcal{J}}{4} \operatorname{Tr}\left[\mathcal{X} F^{2}\right]\right),
\end{equation}
where the $ U(1) $ gauge field  
\fa
F_{\mu \nu} \equiv \nabla_{\mu} A_{\nu}-\nabla_{\nu} A_{\mu} ,
\ffa
and
\begin{equation}
	 X \equiv \operatorname{Tr}[\mathcal{X}], \quad  \mathcal{X}_{\nu}^{\mu}=\frac{1}{2}  \partial^{\mu} \phi \partial_{\nu} \phi, \quad  \operatorname{Tr}\left[\mathcal{X} F^{2}\right] \equiv \mathcal{X}_{\nu}^{\mu} F_{\rho}^{\nu} F_{\mu}^{\rho}.
\end{equation}
We set the $ \Lambda = -1 $ for convenience, implying a normalized $ AdS $ radius. The simplest linear axion model is therefore obtained by setting $ \mathcal{J}  = 0$.

From the action above, the covariant form of the equations of motion are given by
\fa
&&
R_{\mu \nu}-\frac{1}{2} g_{\mu \nu} R-\frac{1}{2} \nabla_{\mu} \phi \nabla_{\nu} \phi-\frac{1}{2}\left(2-\frac{1}{2} \nabla_{\sigma} \phi \nabla^{\sigma} \phi\right) g_{\mu \nu} 
\nonumber
\\
&&
=\frac{1}{2}\left(F_{\mu}^{\sigma} F_{\nu \sigma}-\frac{1}{4} g_{\mu \nu} F_{\rho \sigma} F^{\rho \sigma}\right)+\frac{\mathcal{J}}{4}\left(\frac{1}{2} \nabla_{(\mu \mid} \phi \nabla_{\sigma} \phi\left(F^{2}\right)_{\mid \nu)}^{\sigma}\right. 
\nonumber
\\
&&
\left.+F_{(\mu \mid \sigma}(F \mathcal{X})_{\mid \nu)}^{\sigma}+F_{(\mu \mid \sigma}(\mathcal{X} F)_{\mid \nu)}^{\sigma}-\frac{1}{2} g_{\mu \nu} \operatorname{Tr}\left[\mathcal{X} F^{2}\right]\right),
\
\\
&&
\nabla_{\mu}\left[F^{\mu \nu}-\frac{\mathcal{J}}{2}\left((\mathcal{X} F)^{\mu \nu}-(\mathcal{X} F)^{\nu \mu}\right)\right]=0,
\label{cov-Maxwell}
\\
&&
\nabla_{\mu}\left[\nabla^{\mu} \phi+\frac{\mathcal{J}}{4}\left(F^{2}\right)_{\nu}^{\mu} \nabla^{\nu} \phi\right]=0.
\ffa
Then we can construct the following  homogeneous background ansatz,
\fa
d{s^2} = \frac{1}{{{u^2}}}\left[ - f(u)d{t^2} + \frac{1}{{f(u)}}d{u^2} + d{x^2} \right]\,,
\ffa
\fa
A=A_t(u)dt\,, \qquad \phi=k x\,,
\label{bl-br}
\ffa
with $u\in[0, u_h]$ which indicates the AdS radial direction spanning from the boundary to the horizon defined by $f(u_h)=0$. $k$ is a constant characterizing the strength of broken translation which results in the momentum dissipation in the dual boundary field theory, and $x$ denotes the boundary spatial coordinate. For our model, we can obtain an analytical black hole solution
\begin{equation}
	f(u)=1-\frac{u^{2}}{u_{h}^{2}}+\frac{1}{2} u^{2}\left(k^{2}+\mu^{2}\right) \ln\left(\frac{u}{u_{h}}\right), \quad A_{t}=\mu \ln \left(\frac{u_{h}}{u}\right).
\end{equation}
$\mu$ is an integration constant which should be identified as the chemical potential in the boundary field theory. 
As the $4$-dimensional case, the gauge-axion coupling has no effect on the background solution. Therefore the above solution is also the black hole solution of the simplest linear axion model in $3$-dimension, which is still absent in previous works, as far as we know.
If we set $k = 0$, it will reduce to the $3$-dimensional RN-AdS black hole solution \cite{Ren:2010ha,Liu:2011fy}. The Hawking temperature is given by
\fa
&&
T = \frac{{4 - {u_h^2}\left( {{k^2} + {\mu ^2}} \right)}}{{8\pi {u_h}}}\,.
\ffa
Making the following rescaling
\fa
\label{rescalign}
(t,u,x) \rightarrow u_h (t,u,x)\,,\qquad
(k,\mu)\rightarrow\frac{1}{u_h}(k,\mu)\,,\quad
\ffa
we can set $u_h=1$. It is easy to find that this above black hole solution is determined by two dimensionless parameters $\hat{T}\equiv T/\mu$ and $\hat{k}\equiv k/\mu$.
\section{Thermoelectric conductivities} \label{CC}

The thermoelectric transport properties in the dual field theory can be captured by the generalized Ohm's law
\begin{eqnarray}\label{ohm}
	\left(  \begin{array}{c}
		J \\
		Q \\
	\end{array}\right)=\left(
	\begin{array}{cc}
		\sigma & \alpha \,T \\
		\bar{\alpha}\, T & \bar{\kappa} \,T \\
	\end{array}
	\right)
	\left(  \begin{array}{c}
		E \\
		-\nabla T/T \\
	\end{array}\right).
\end{eqnarray}
$J$ and $Q$ are electric and thermal currents respectively, which are generated under an external electric field $E$ and a thermal gradient $\xi\equiv-\nabla T/T$. The corresponding transport coefficients are the so-called electric conductivity $\sigma$,  thermoelectric conductivities $\alpha$ and $\bar{\alpha}$, and thermal conductivity $\bar{\kappa}$. We will compute these quantities analytically from holography via a simple method called ``membrane paradigm'' \cite{Iqbal:2008by,Blake:2013bqa,Donos:2014uba,Donos:2014cya,Baggioli:2016pia,Baggioli:2017ojd}. The key point of this method is to construct a radially conserved current that one can read off data from the black hole horizon directly.

To this end, we will follow the procedure proposed by \cite{Donos:2014cya} and consider the following consistent perturbations
\fa
&&
\delta {{\rm{g}}_{tx}} = \frac{1}{u^2}\left( - \xi f(u) t + h_{tx}(u)\right),\qquad
\delta {g_{ux}} = \frac{{{h _{ux}(u)}}}{{{u^2}}},\nonumber\\
\label{eq:8}
&&
\delta {A_x} = \left( {\xi {A_t(u)} - {E}} \right)t + {a_x(u)},\qquad
\delta {\phi^x} = \psi(u).
\label{eq:9}
\ffa
Then the linearized equations of the metric field, scalar field and gauge field are given by
\begin{align}
&f\left(u^{2}\left(-4+\mathcal{J} k^{2} u^{2}\right) \mu a_{x}^{\prime}+4 t \xi f^{\prime}-4 h_{t x}^{\prime}+4 u\left(t \xi\left(k^{2}+\mu^{2}-f^{\prime \prime}\right)+h_{t x}^{\prime \prime}\right)\right)  \nonumber \\
&-k^{2} u\left(4+\mathcal{J} u^{2} \mu^{2}\right) h_{t x}=0,\\
&u\left( { - 4 + {\cal J}{k^2}{u^2}} \right)\mu \left( {E + \mu \xi \ln u} \right) + 4\xi f' + k(4 + {\cal J}{u^2}{\mu ^2})f(k{h_{ux}} - \psi ') = 0,\label{eq12}\\
&ku(4\xi  + {\cal J}{u^2}\mu (2E + \mu \xi ) + 2{\cal J}{u^2}{\mu ^2}\xi \ln u) + u\left( {4 + {\cal J}{u^2}{\mu ^2}} \right){f^\prime }\left( {k{h_{ux}} - {\psi ^\prime }} \right)\nonumber \\
&+f\left(k\left(-4+\mathcal{J} u^{2} \mu^{2}\right) h_{u x}+4 \psi^{\prime}+u(k\left(4+\mathcal{J} u^{2} \mu^{2}\right) h_{u x}^{\prime}-4 \psi^{\prime \prime}-\mathcal{J} u \mu^{2}t(\psi^{\prime}+u \psi^{\prime \prime}))\right)=0,\\
&-2 \mathcal{J} k^{2} u \mu h_{t x}+\left(-4+\mathcal{J} k^{2} u^{2}\right)\left(u a_{x}^{\prime} f^{\prime}-\mu h_{t x}^{\prime}\right)+f\left(-4+3 \mathcal{J} k^{2} u^{2}+a_{x}^{\prime}+u\left(-4+\mathcal{J} k^{2} u^{2}\right) a_{x}^{\prime \prime}\right)=0.\label{maxwell1}
\end{align}

The key point of this method is to construct the radially conserved currents in the bulk. In our model, they are given by
\fa
&&
J = \frac{1}{4}\left( { - 4 + \mathcal{J}{k^2}{u^2}} \right)\left( { - \mu {h_{tx}} + uf{a_x}^\prime } \right),
\
\\
&&
Q=\frac{f^2}{u}\left(\frac{h_{tx}}{f}\right)'-A_t J.
\ffa
The electric current $J$ can also be directly read off from the covariant Maxwell equation \eqref{cov-Maxwell}. Also, we have checked that $\partial_u Q=0$ and so the thermal current $Q$ constructed above is indeed a radially conserved quantity in the bulk.

Given the above conserved currents, the DC transports can be determined by the regularity of the perturbation variables at the horizon, which gives
\begin{equation}
h_{tx}\approx f h_{ux}+\cdots,\qquad a_x'\approx -\frac{E}{f}+\cdots,
\end{equation}
along with the following constraint derived from Eq.\eqref{eq12}
\begin{equation}
	h_{u x}= - \frac{{u\left( { - 4 + \mathcal{J}{k^2}{u^2}} \right)\mu \left( {E + \mu \xi \ln u} \right) + 4\xi f'}}{{{k^2}\left( {4 + \mathcal{J}{u^2}{\mu ^2}} \right)f}}.
	\label{constraint}
\end{equation}

Collecting all the above information, we write down the expression of the conserved currents
\begin{align}
	&J = - \frac{{\left( { - 4 + \mathcal{J}{k^2}{u^2}} \right)\left( {4 + \mathcal{J}{\mu ^2}} \right)\left( { - 4Eu\left( {{k^2} + {\mu ^2}} \right) + u\left( { - 4 + \mathcal{J}{k^2}{u^2}} \right){\mu ^3}\xi \ln u + 4\mu \xi f'} \right)}}{{4{k^2}\left( { - 4 + \mathcal{J}{k^2}} \right)\left( {4 + \mathcal{J}{u^2}{\mu ^2}} \right)}},
\end{align}	
\begin{align}
	&Q =  - \frac{1}{{4{k^2}u{{\left( {{\cal J}{\mu ^2}{u^2} + 4} \right)}^2}}}(4f\left( {{\cal J}{\mu ^2}{u^2} + 4} \right)\nonumber\\
	&\left( {\mu \left( {E\left( {3{\cal J}{k^2}{u^2} - 4} \right) + \mu \xi \left( {{\cal J}{k^2}{u^2} - 4} \right) + \mu \xi \ln u\left( {3{\cal J}{k^2}{u^2} - 4} \right)} \right) + 4\xi f''} \right) -\nonumber \\
	&4\left( {{\cal J}{\mu ^2}u\left( {uf' + 2f} \right) + 4f'} \right)\left( {\mu u\left( {{\cal J}{k^2}{u^2} - 4} \right)(E + \mu \xi \ln u) + 4\xi f'} \right) +\nonumber\\
	&\mu u\ln \frac{1}{u}\left( {4 - {\cal J}{k^2}{u^2}} \right)\left( {{\cal J}{\mu ^2}{u^2} + 4} \right)\left( {4Eu\left( {{k^2} + {\mu ^2}} \right) - 4\mu \xi f' - u{\mu ^3}\xi \ln u\left( {{\cal J}{k^2}{u^2} - 4} \right)} \right)).
\end{align}
From these expressions the DC conductivities follow directly 
\begin{align}
	&\sigma  = \frac{{\partial J}}{{\partial {E}}} = \left( {1 - \frac{1}{4}{\cal J}{k^2} + \frac{{{{\left( {1 - \frac{1}{4}{\cal J}{k^2}} \right)}^2}{\mu ^2}}}{{\left( {1 + \frac{1}{4}{\cal J}{\mu ^2}} \right){k^2}}}} \right),\\
	&\bar \alpha  = \frac{1}{T}\frac{{\partial Q}}{{\partial {E}}} = \frac{{4\pi \mu \left( {1 - \frac{1}{4}{\cal J}{k^2}} \right)}}{{{k^2}\left( {1 + \frac{1}{4}{\cal J}{\mu ^2}} \right)}},\\
	&\alpha  = \frac{1}{T}\frac{{\partial J}}{{\partial \xi }} = \frac{{4\pi \mu \left( {1 - \frac{1}{4}{\cal J}{k^2}} \right)}}{{{k^2}\left( {1 + \frac{1}{4}{\cal J}{\mu ^2}} \right)}},\\
	&\bar \kappa  = \frac{1}{T}\frac{{\partial Q}}{{\partial \xi }} = \frac{{16{\pi ^2}T}}{{{k^2}\left( {1 + \frac{1}{4}{\cal J}{\mu ^2}} \right)}}\label{DC1v0}.
\end{align}
Besides the thermal conductivity $\bar \kappa$ at zero electric field, a more readily measurable quantity is the thermal conductivity $\kappa$ at zero electric current defined by $\kappa=\bar \kappa-\bar \alpha \alpha T/\sigma$. It can be explicitly expressed as
\begin{equation}
	\kappa  = \frac{{16{\pi ^2}T}}{{\left( {{k^2} + {\mu ^2}} \right)}}.
	\label{kappav0}
\end{equation}

We are interested in the dimensionless DC thermoelectric conductivities defined by $\hat{\sigma}\equiv\sigma\mu$, $\hat{\alpha}\equiv\alpha\mu$ and $\hat{\bar{\alpha}}\equiv\bar{\alpha}\mu$. Notice that $\bar{\kappa}$ and $\kappa$ themselves are dimensionless. In terms of the dimensionless parameters $\hat{T}$ and $\hat{k}$, we re-express the dimensionless DC thermoelectric conductivities as what follows
\begin{align}
	&\hat \sigma=\frac{{\mu ( {1 + {{\hat k}^2}} )\left( {1 - \frac{1}{4}\mathcal{J}{{\hat k}^2}{\mu ^2}} \right)}}{{{{\hat k}^2}( {1 + \frac{1}{4}\mathcal{J}{\mu ^2}} )}}\,,\nonumber\\
	&\hat{\alpha}=   \frac{{4\pi {\mu ^2}( {1 - \frac{1}{4}{{\hat k}^2}\mathcal{J}} )}}{{{{\hat k}^2}( {1 + \frac{1}{4}\mathcal{J}{\mu ^2}} )}}\,,\,\,\,\,\,\,\,\,\,\,\,\,
	\hat{\bar \alpha}= \frac{{4\pi {\mu ^2}( {1 - \frac{1}{4}{{\hat k}^2}\mathcal{J}} )}}{{{{\hat k}^2}( {1 + \frac{1}{4}\mathcal{J}{\mu ^2}} )}}\,,\nonumber\\
	&\bar{\kappa}= \frac{{16{\pi ^2}\hat T}}{{{{\hat k}^2}\mu \left( {1 + \frac{1}{4}\mathcal{J}{\mu ^2}} \right)}}\,,\,\,\,\,\,\,\,\,\,\,\,\,\,
	\kappa=\frac{{16{\pi ^2}\hat T}}{{( {1 + {{\hat k}^2}} )\mu }}\,.
	\label{kappa}
\end{align}

At zero temperature, $\mu=2/\sqrt{1+\hat{k}^2}$, for which the DC electric conductivity reduces to 
\begin{equation}
	\label{sigmavT0}
	\hat{\sigma}_0=\frac{2\sqrt{1+\hat{k}^2}}{\hat{k}^2}\Big(\frac{1+(1-\mathcal{J})\hat{k}^2}{1+\mathcal{J}+\hat{k}^2}\Big)\,.
\end{equation}
It is easy to see that if the coupling parameter $\mathcal{J}$ satisfies\footnote{In Appendix \ref{appendix-A}, we examine the spectrum of quasi-normal modes (QNMs) of the vector mode at zero charge density and find that they are all in the lower half of the complex frequency plane for $-1\leq\mathcal{J}\leq 1$. It indicates that this system at zero charge density is stability under the perturbation of vector mode when the gauge-axion coupling belongs to $-1\leq\mathcal{J}\leq 1$. Also, we inspect the graviton mass of vector mode at finite charge density and find that it is real for $-1\leq\mathcal{J}\leq 1$.}
\begin{equation}
	\label{Jregion}
	-1\leq\mathcal{J}\leq 1\,,
\end{equation}
the DC electric conductivity at zero temperature $\hat{\sigma}_0$ holds positive. However, once $\mathcal{J}$ is beyond the above region, $\hat{\sigma}_0$ will inevitably become negative\footnote{For $\mathcal{J}>1$, $\hat{\sigma}_0$ is negative provided $\hat{k}>1/(J-1)$. While for $\mathcal{J}<1$, $\hat{\sigma}_0$ becomes negative if $\hat{k}$ is small enough.}. Through this paper, we shall constrain $\mathcal{J}$ in the region of $-1\leq\mathcal{J}\leq 1$.

At both zero temperature and the strong momentum dissipation limit ($\hat{k} \rightarrow +\infty$), $\hat{\sigma}_0$ can be expanded as up to the first order
\begin{equation}
	\label{sigma0infty}
	\hat{\sigma}_0\underset{\hat{k} \rightarrow+\infty}{\simeq} \frac{2(1-\mathcal{J})}{\hat{k}}\,.
\end{equation}
It indicates that at both zero temperature and the strong momentum dissipation limit, $\hat{\sigma}_0$ asymptotes to zero. In addition, for fixed $\hat{k}$, $\hat{\sigma}_0$ decreases with $\mathcal{J}$ approaching $1$ and finally vanishes when $\mathcal{J}=1$.

We also expand $\hat{\sigma}_0$ at both zero temperature and the weak momentum dissipation limit ($\hat{k}\rightarrow 0$) up to the first order, which is given as
\begin{equation}
	\label{sigma00}
	\hat{\sigma}_0\underset{\hat{k} \rightarrow 0}{\simeq} \frac{2}{(1+\mathcal{J})\hat{k}^2}\,.
\end{equation}
It is obvious that as the momentum dissipation becomes weak, $\hat{\sigma}_0$ becomes large and finally asymptotes to infinity in the limit of $\hat{k}\rightarrow 0$. In addition, we also observe that for fixed $\hat{k}$, $\hat{\sigma}_0$ becomes large with $\mathcal{J}$ approaching $-1$ and finally asymptotes to infinity when $\mathcal{J}\rightarrow -1$. Finally, from the expressions of $\hat{\sigma}_0$ at the strong/weak momentum dissipation limit, it is also easy to confirm the bound of $\mathcal{J}$ (Eq.\eqref{Jregion}).

\begin{figure}[H]
	\begin{minipage}[t]{0.45\linewidth}
		\centering
		\includegraphics[width=2.8in]{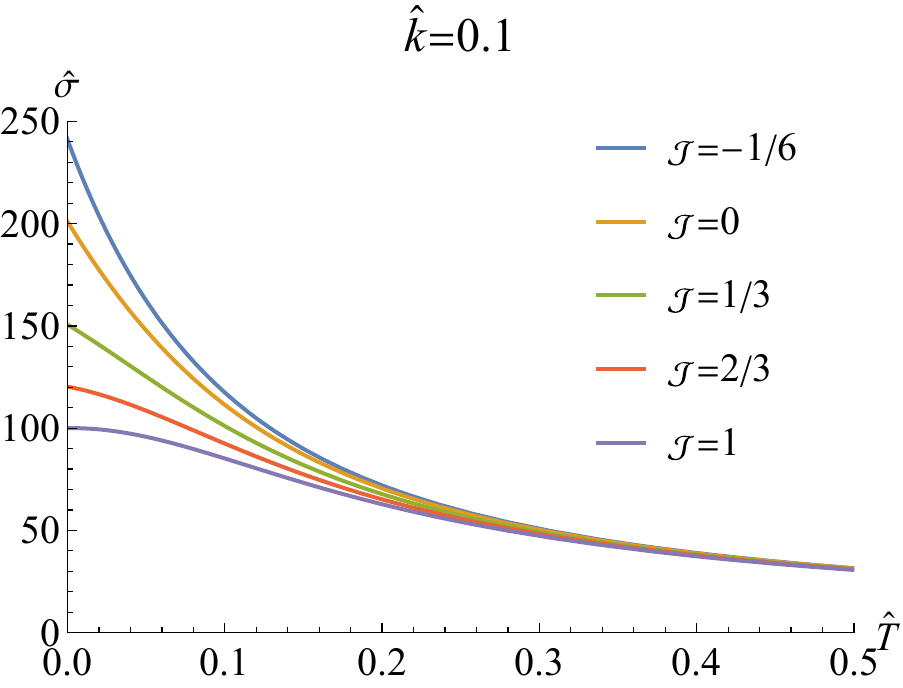}
	\end{minipage}
	\begin{minipage}[t]{0.5\linewidth}
		\centering
		\includegraphics[width=2.8in]{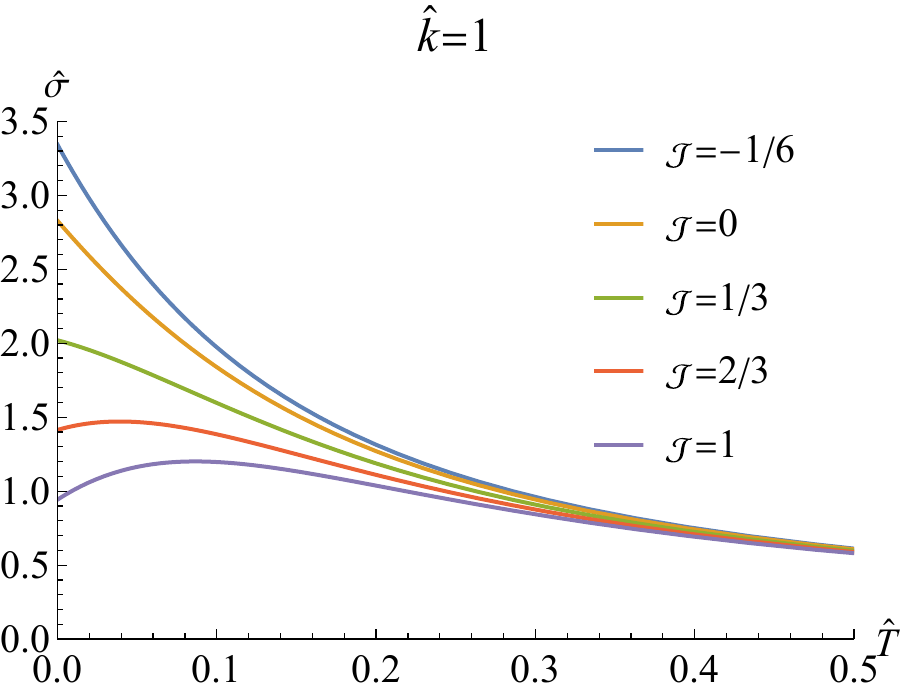}
	\end{minipage}
	\caption{DC electric conductivity $\hat{\sigma}$ as the function of $\hat{T}$ with different $\mathcal{J}$ and $\hat{k}$.}
	\label{fig:2}
\end{figure}
We are also interested in the temperature dependent DC electric conductivity $\hat{\sigma}$. Fig.\ref{fig:2} shows $\hat{\sigma}$ as the function of $\hat{T}$ with different $\mathcal{J}$ and $\hat{k}$. We find that $\hat{\sigma}$ is also temperature dependent even when $\mathcal{J}=0$, for which this model reduces to the usual 3-dimensional holographic axion model.
It is a novel property different from that of the usual $4$-dimensional axion model whose DC electric conductivity is temperature independent \cite{Andrade:2013gsa}.

\begin{figure}[H]
	\centering
	\includegraphics[width=0.4\textwidth]{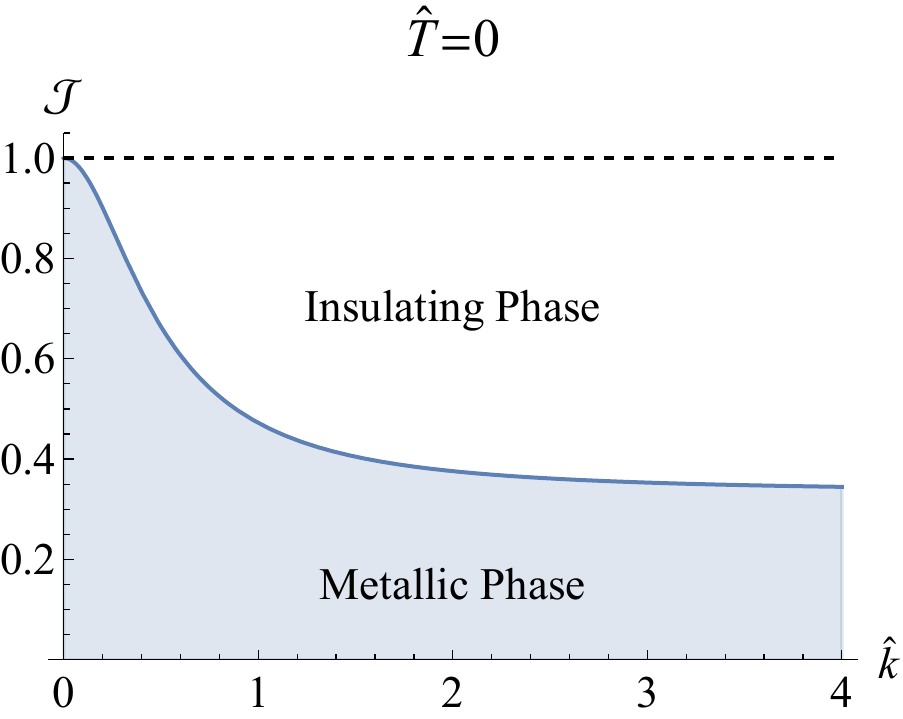}
	\caption{Phase diagram over $\{\hat{k},\mathcal{J}\}$ at zero temperature. The blue
		line is the critical line of the phase transition.}
	\label{fig:7}
\end{figure}

When we turn on the gauge-axion coupling $\mathcal{J}$, $\hat{\sigma}$ decreases with $\mathcal{J}$ increasing at zero temperature. 
As the temperature increases, $\hat{\sigma}$ tends to the same constant value. 
For all negative $\mathcal{J}$ and small positive $\mathcal{J}$, this holographic system exhibits metallic behavior (Fig.\ref{fig:2}). However, we observe that for large $\mathcal{J}$ and certain $\hat{k}$, $\hat{\sigma}$ as the function of $\hat{T}$ may be non-monotonic (Fig.\ref{fig:2}). It indicates that there is insulating behavior for certain $\hat{k}$ and $\mathcal{J}$.
Notice that here we adopt the operational definition as most of holographic references \cite{Baggioli:2016rdj,Donos:2012js,Donos:2013eha,Donos:2014uba,Ling:2015ghh,Ling:2015dma,Ling:2015epa,Ling:2015exa,Ling:2016wyr,Ling:2016dck,Baggioli:2014roa,Baggioli:2016oqk,Baggioli:2016oju,Donos:2014oha,Liu:2021stu}, to identify the metallic phase and insulating phase, i.e., $\partial_{\hat{T}}\hat{\sigma}<0$ indicating metallic phase and $\partial_{\hat{T}}\hat{\sigma}>0$ meaning insulating phase. The critical point (line) can be determined by $\partial_{\hat{T}}\hat{\sigma}=0$. Using this definition, we show the phase diagram $\{\hat{k},\mathcal{J}\}$ at zero temperature in Fig.\ref{fig:7}. 
We find that when $\mathcal{J}$ is small, the system is metallic for all $\hat{k}$. However, when $\mathcal{J}$ is large, there is a MIT that the system changes from metallic phase to insulating phase with $\hat{k}$ increasing. 
It is different from that of the $4$-dimensional case, where for fixed gauge-axion coupling, we cannot observe a MIT when changing the strength of momentum dissipation \cite{Gouteraux:2016wxj,Baggioli:2016pia,Li:2018vrz,Liu:2022bam}.

\begin{figure}[H]
	\begin{minipage}[t]{0.5\linewidth}
		\centering
		\includegraphics[width=2.8in]{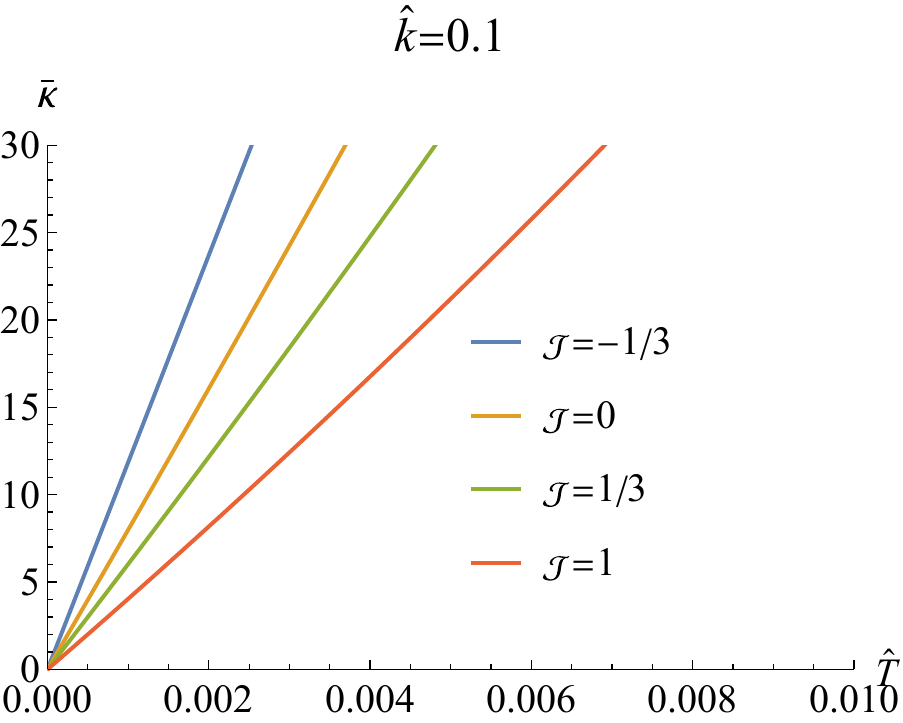}
	\end{minipage}
	\begin{minipage}[t]{0.5\linewidth}
		\centering
		\includegraphics[width=2.8in]{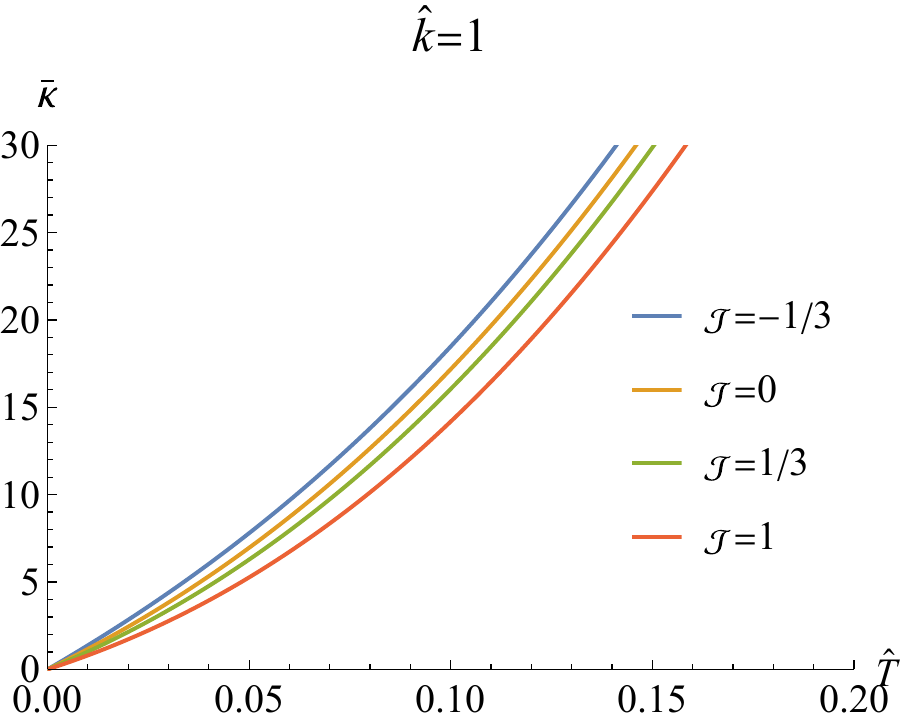}
	\end{minipage}
	\caption{Thermal conductivity $\bar{\kappa}$ at zero electric field as the function of $\hat{T}$ with different  $\mathcal{J}$ and $\hat{k}$.}
	\label{fig:3}
\end{figure}
\begin{figure}[H]
	\centering
	\includegraphics[width=0.4\textwidth]{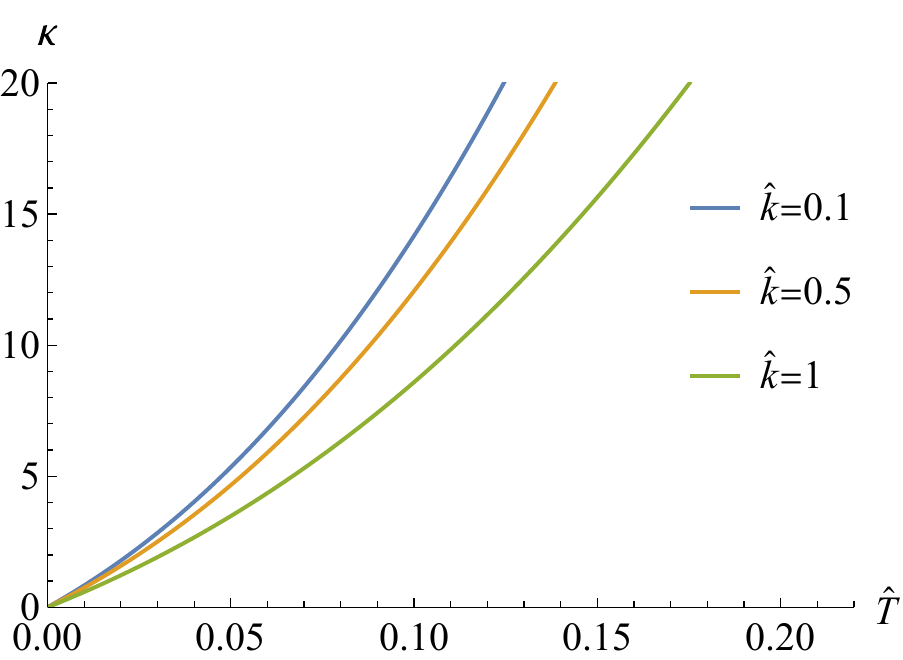}
	\caption{Thermal conductivity $\kappa$ at zero electric current as the function of $\hat{T}$ with different  $\hat{k}$.}
	\label{fig:5}
\end{figure}

Further, we show the thermal conductivity $\bar{\kappa}$ at zero electric field as the function of $\hat{T}$ with different $\mathcal{J}$ and $\hat{k}$ in Fig.\ref{fig:3}. We see that $\bar{\kappa}$ decreases with the temperature for different $\mathcal{J}$ and $\hat{k}$, and finally it declines to zero as the temperature drops to zero. At finite temperature, $\bar{\kappa}$ is suppressed with the gauge-axion coupling increasing. Notice that the thermal conductivity $\kappa$ at zero electric current is independent of the gauge-axion coupling, which can be seen from Eq.\eqref{kappa}. The temperature-dependent behavior of $\kappa$ is closely similar to that of $\bar{\kappa}$. That is to say, $\kappa$ decreases with the temperature and drops to zero at zero temperature (see Fig.\ref{fig:5}). Therefore, the $3$-dimensional holographic system studied here is also an ideal thermal insulator, which is similar to the higher dimensional holographic system \cite{Gouteraux:2016wxj,Baggioli:2016pia,Li:2018vrz}.

\begin{figure}[H]
	\begin{minipage}[t]{0.5\linewidth}
		\centering
		\includegraphics[width=2.8in]{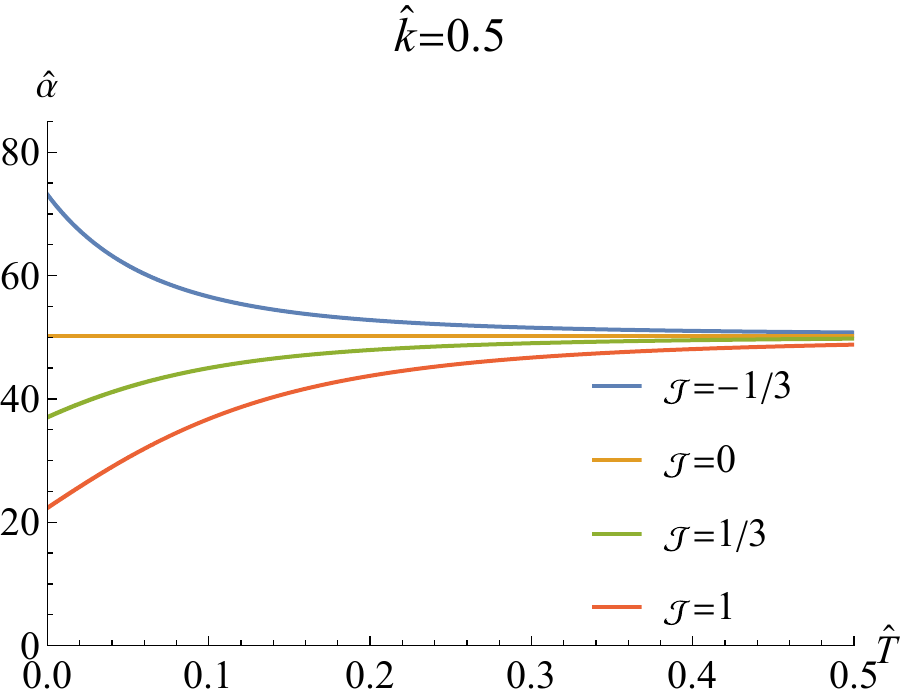}
	\end{minipage}
	\begin{minipage}[t]{0.5\linewidth}
		\centering
		\includegraphics[width=2.8in]{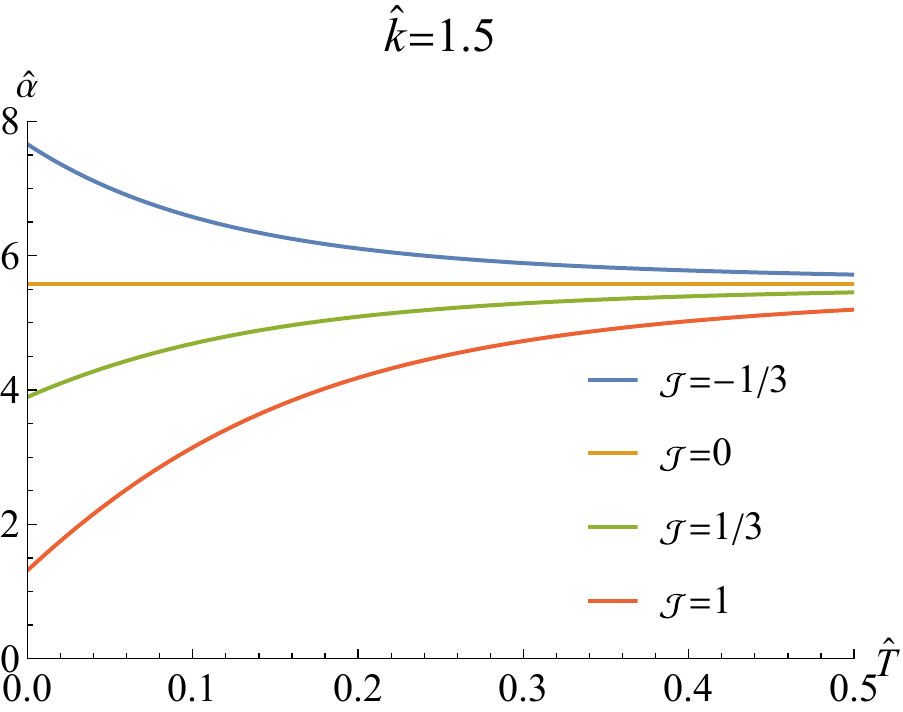}
	\end{minipage}
	\caption{Thermoelectric conductivity $\hat{\alpha}$ as the function of $\hat{T}$ with different  $\mathcal{J}$   and $\hat{k}$.}
	\label{fig:4}
\end{figure}
Then, we also study the behavior of thermoelectric conductivity $\hat{\alpha}$, which is shown in Fig.\ref{fig:4}. From this figure, we clearly see that the thermoelectric conductivity is independent of the temperature for $ \mathcal{J} =0 $, which is different from the case of $4$-dimensional holographic system \cite{Gouteraux:2016wxj,Baggioli:2016pia,Li:2018vrz}. However, once the gauge-axion coupling is turned on, as the temperature rises the thermoelectric conductivity increases for $\mathcal{J}>0$ but decreases for $\mathcal{J}<0$.

We also expand $\hat{\alpha}$ up to the first order at both zero temperature and the strong momentum dissipation limit, which is given as
\begin{equation}
	\hat{\alpha}\underset{\hat{k} \rightarrow +\infty}{\simeq}\frac{4\pi(1-\mathcal{J})}{\hat{k}^2}\,.
\end{equation}
It is obvious that in the region of $-1\leq\mathcal{J}\leq 1$, $\hat{\alpha}$ holds positive. In addition, from the above equation, it is also easy to find that for fixed $\mathcal{J}$, $\hat{\alpha}$ decreases with $\hat{k}$ increasing at zero temperature. It is consistent with the result shown in Fig.\ref{fig:4}.

\begin{figure}[H]
	\begin{minipage}[t]{0.5\linewidth}
		\centering
		\includegraphics[width=3.0in]{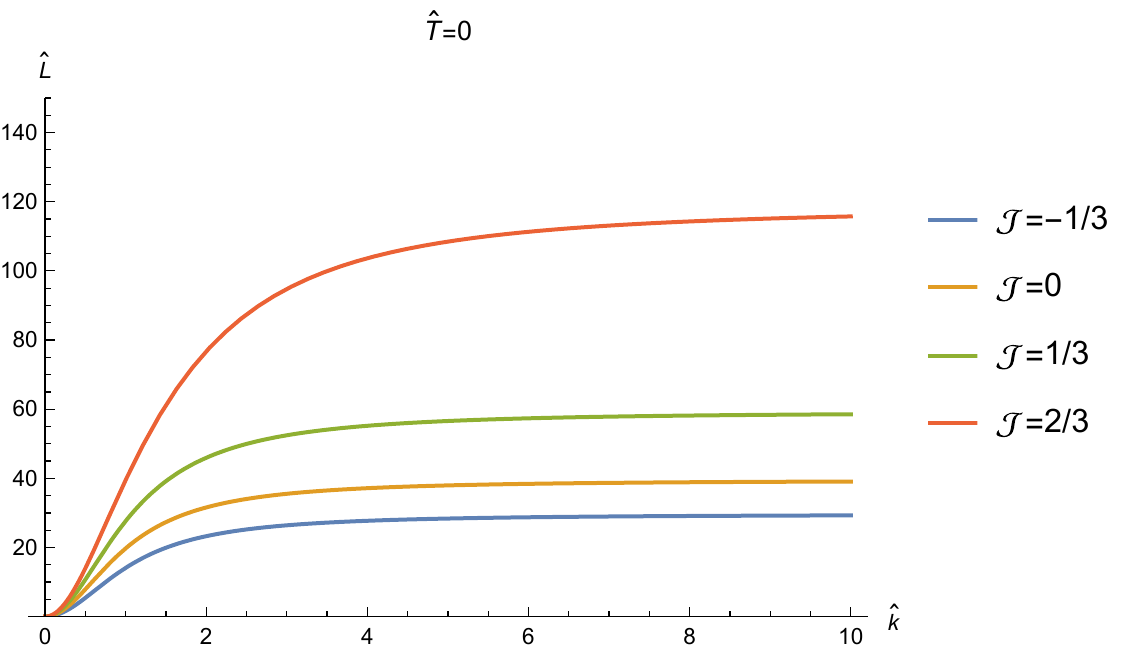}
	\end{minipage}
	\begin{minipage}[t]{0.5\linewidth}
		\centering
		\includegraphics[width=3.0in]{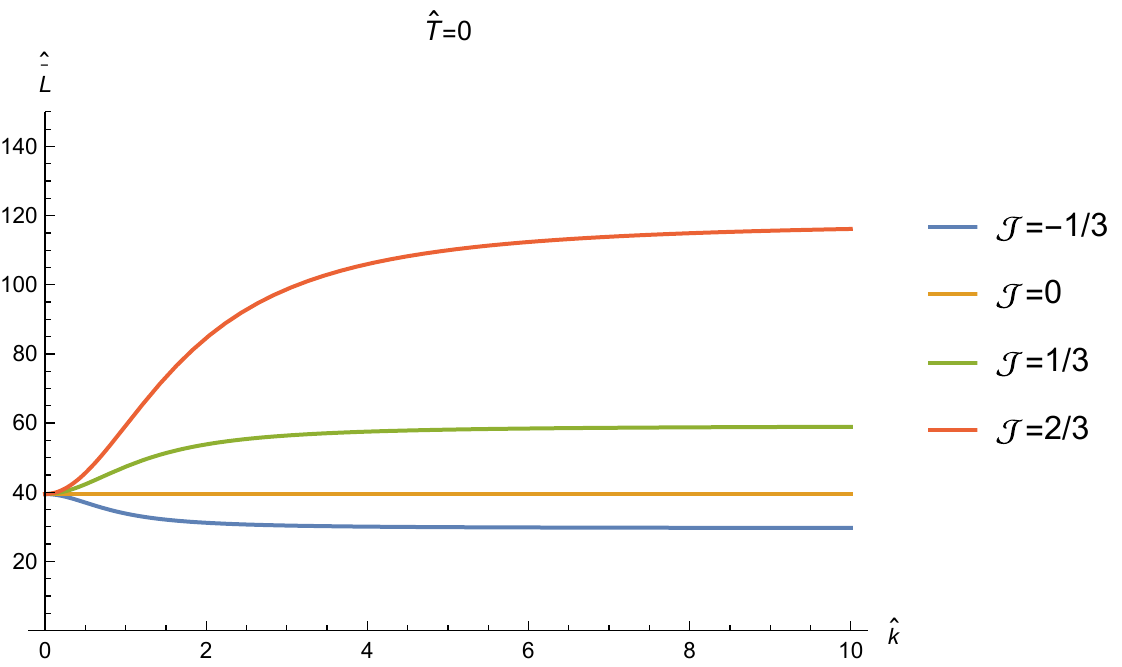}
	\end{minipage}
	\caption{The Lorentz ratios $\hat{L}$ and $\hat{\bar{L}}$ as the function of $\hat{k}$ for different $\mathcal{J}$ at $\hat{T}=0$.}
	\label{fig:6}
\end{figure}

Finally, we briefly check the Wiedemann-Franz (WF) law. It states that the ratio of the thermal conductivity to the electric conductivity, also called Lorentz ratio, is a constant for Fermi liquid \cite{ziman2001electrons}. For the strongly
interacting non-Fermi liquids, the WF law is violated \cite{Mahajan:2013cja}.  As far as we know, the WF law is also violated in holographic dual system, see e.g., \cite{Kuang:2017rpx,Kim:2014bza,Donos:2014cya,Wu:2018zdc}.
In our model, it is easy to derive the Lorentz ratios, which are
\begin{align}
	&\hat{L}\equiv \frac{{\kappa }}{{\hat \sigma \hat T}} = \frac{{16{\hat{k}^2}{\pi ^2}(1 + \frac{1}{4}{\cal J}{\mu ^2})}}{{\mu^2(1 + {{\hat k}^2})^2(1 - \frac{1}{4}{\cal J}\hat{k}^2{\mu ^2})}},\\
	&\hat{\bar L }\equiv\frac{\bar \kappa}{\hat{\sigma}\hat{T}}=\frac{16 \pi ^2}{\mu ^2 (1+\hat{k}^2)  (1-\frac{1}{4} \mathcal{J}
		\hat{k}^2 \mu ^2)}.
\end{align}
It is obvious that the Lorentz ratios depend on the strength of momentum dissipation and the gauge-axion coupling (see Fig.\ref{fig:6} for the case of $\hat{T}=0$). This means that the WF law is also violated in $3$-dimensional holographic effective theory with gauge-axion coupling.

\section{Conclusion and discussion}\label{conclu}

In this paper, we construct a $3$-dimensional holographic effective theory with gauge-axion coupling. The black hole solution is analytically worked out. As the $4$-dimensional case, the gauge-axion coupling has no effect on the background solution. Therefore this black hole solution is also the one of the simplest $3$-dimensional axion model.

Then we study the DC thermoelectric conductivities. A novel property is that DC electric conductivity is temperature dependent even for vanishing gauge-axion coupling, which is different from that of $4$-dimensional axion model whose DC electric conductivity is temperature independent.
In particular, the introduction of the gauge-axion coupling induces a MIT at zero temperature.

Further, the properties of other DC thermoelectric conductivities are also discussed. Finally, we also check the WF law and find that similarly to $4$-dimensional holographic model, the WF law is violated in our $3$-dimensional holographic system.

This work is the first step towards exploring the $3$-dimensional black hole solution of holographic model with momentum dissipation and the corresponding transport properties. It will be also interesting to further study the Alternating current (AC) conductivity of our model. And then, we will also extend our study in holographic Q-lattice model, which has been well studied in $4$-dimensional holographic systems. It will be surely interesting to investigate MIT in $3$-dimensional holographic system.
More importantly, we expect that some peculiar properties emerging in $3$-dimensional holographic system, which is different from $4$-dimensional case, such that we can address some exclusive characteristics in $(1+1)$ dimensional condensed matter system and explore the basic principle hidden in them.

\begin{acknowledgments}
	
We are very grateful to Wei-Jia Li, Peng Liu and Dan Zhang for helpful discussions and suggestions. This work is supported by the Natural Science Foundation of China under Grants Nos. 11775036, 12147209, the Postgraduate Research \& Practice Innovation Program of Jiangsu Province under Grant No. KYCX20\_2973 and KYCX21\_3192, and Top Talent Support Program from Yangzhou University.
\end{acknowledgments}

\appendix
\section{Bounds on the coupling}\label{appendix-A}

The requirement of positive definiteness of DC conductivity gives the constraint of $\mathcal{J}$ in the region $-1\leq\mathcal{J}\leq 1$ (Eq.\eqref{Jregion}). In this appendix, we shall examine in the region of $-1\leq\mathcal{J}\leq 1$, the spectrum of QNMs of the vector mode at zero charge density to see the stability of the system under the perturbation of vector mode. Also, we shall inspect the graviton mass of vector mode at finite charge density to see if it is real for $-1\leq\mathcal{J}\leq 1$.

\subsection{QNMs at zero charge density}

At zero charge density, the Hawking temperature reduces to
\fa
&&
T=\frac{4-k^2}{8\pi}
\,.
\label{T-SS}
\ffa
Therefore, the system is determined by one dimensionless parameter $\tilde{k}=k /\mathfrak{p}$ with $\mathfrak{p}\equiv 8\pi T$. The $a_x$ perturbation decouples from the other perturbations. Decomposing the $a_x$ perturbation in the Fourier space as $a_x(t,u)\sim e^{-i\omega t} a_x(u)$, the perturbative equation can be evaluated as
\begin{equation}
	\label{eq:axeomJmu=0}
\left[\mathcal{D}fa_x'\right]'+\frac{\mathcal{D}}{f} \omega^2 a_x=0\,,
\end{equation}
where
\begin{equation}
\mathcal{D}=u(\mathcal{J} k^2 u^2-4)\,.
\end{equation}
After making a transformation
\begin{equation}
\frac{dz}{du}=\frac{1}{f(u)}\,,
\end{equation}
we can formulate this perturbation equation in the ingoing Eddington coordinate, which is given as (see \cite{Horowitz:1999jd,Jansen:2017oag,Wu:2018vlj,Fu:2018yqx,Liu:2020lwc} for more details)  
\begin{equation}
	a_x''(u)+\left(\frac{4-3\mathcal{J}k^2u^2}{u(4-\mathcal{J} k^2 u^2)}+\frac{2i \omega+f'(u)}{f(u)}\right)a_x'(u)+\left(\frac{i\omega(3\mathcal{J}k^2u^2-4)}{u(\mathcal{J}k^2u^2-4)f(u)}\right)a_x(u)=0\,.
\end{equation}
We shall use the pseudospectral methods outlined in \cite{Jansen:2017oag} to solve the above perturbative equation and obtain the spectrum of QNMs.

\begin{figure}[H]
	\centering
	\includegraphics[width=0.55\textwidth]{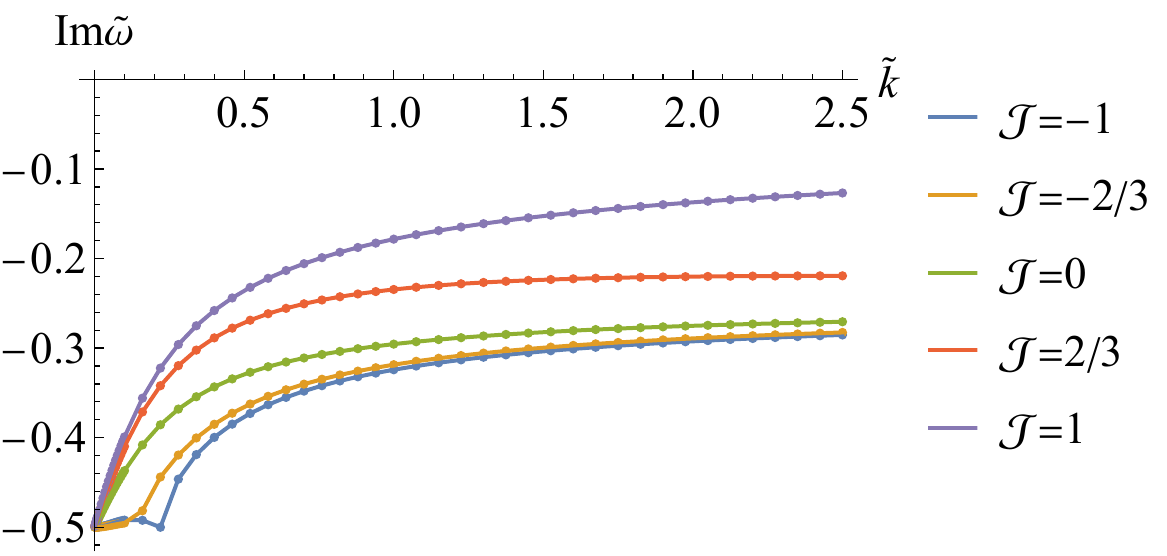}
	\caption{The imaginary parts of spectrum of QNMs with respect to $\tilde{k}$ for various $\mathcal{J}$. Here $\tilde{\omega}\equiv\omega/\mathfrak{p}$ is the dimensionless frequency.}
	\label{fig:QNMs}
\end{figure}

Fig.\ref{fig:QNMs} shows the spectrum of QNMs (fundamental modes) with respect to $\tilde{k}$ for various $\mathcal{J}$ belonging to $-1\leq\mathcal{J}\leq 1$. We find that the imaginary parts of spectrum of QNMs are all negative. It means that the system at zero charge density is stable under the $a_x$ perturbation for $-1\leq\mathcal{J}\leq 1$.

At finite charge density, the perturbation equations are coupled. Especially, for 3-dimensional holographic system, if we package the perturbation equations into the Schr\"odinger form, the Schr\"odinger potential diverges as $1/u^2$ at the UV boundary ($u\rightarrow 0$).
Therefore, it is quite hard to numerically work out the spectrum of QNMs and beyond the scope of this paper. We shall come back this issue in future. Alternatively, we shall examine the graviton mass of vector mode at finite charge density to see if it is real for $-1\leq\mathcal{J}\leq 1$ in next subsection.

\subsection{Graviton mass of vector mode}

In our model, the effective graviton mass should be read as \cite{Gouteraux:2016wxj,Li:2018vrz}
\fa
	\mathcal{M}^2_g(u)=k^2\left(1+\frac{1}{4}\mathcal{J}\mu^2u^2\right),
	\label{mass0}
\ffa 
which depends on the radial direction. Obviously, $\mathcal{M}^2_g(u)>0$ at zero charge density.

\begin{figure}[H]
	\begin{minipage}[t]{0.5\linewidth}
		\centering
		\includegraphics[width=2.8in]{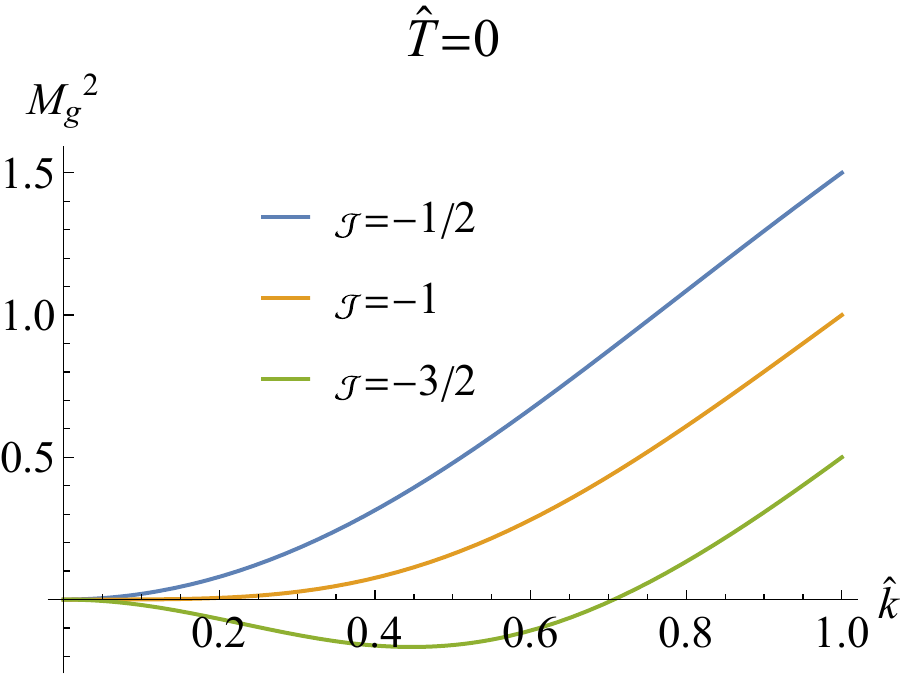}
	\end{minipage}
	\begin{minipage}[t]{0.5\linewidth}
		\centering
		\includegraphics[width=2.8in]{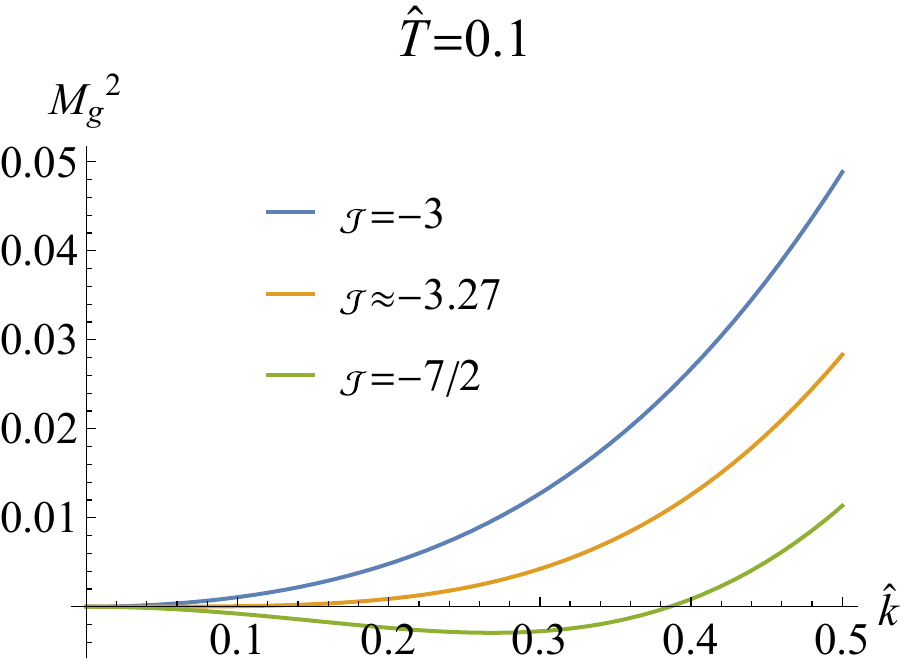}
	\end{minipage}
	\caption{The effective graviton mass at the horizon for the zero and finite temperature respectively.}
	\label{mass}
\end{figure}
It is enough to analyze the graviton mass at the horizon $(u=1)$.
We start by analyzing the mass at zero temperature, for which it takes the simple form
\begin{equation}
	\mathcal{M}_g^2=\frac{4\hat{k}^2(1+\mathcal{J}+\hat{k}^2)}{(1+\hat{k}^2)^2}\approx4(1+\mathcal{J})\hat{k}^2+\mathcal{O}(\hat{k}^3),
\end{equation}
where the latter formula is the expansion for small $\hat{k}$. For $\mathcal{J}\geq-1$, the mass is always positive. However for $\mathcal{J}<-1$, the mass at small enough $\hat{k}$ is negative. This is depicted in the left plot of Fig.\ref{mass}. When turning on the temperature, for example $\hat{T}=0.1$, it is easy to show that the allowed range of the coupling $\mathcal{J}$ for the requirement of the positive mass is $\mathcal{J}\gtrsim -3.27$ (right plot in Fig.\ref{mass}), which is larger than the result at zero temperature. Thus we obtain the lower bound of the coupling $\mathcal{J}\geq-1$ by requiring effective graviton mass to be positive in general. In other words, the graviton mass of our model at finite charge density is real for $-1\leq\mathcal{J}\leq 1$.

\bibliographystyle{style1}
\bibliography{ref}

\providecommand{\href}[2]{#2}\begingroup\raggedright\begin{thebibliography}{10}

\bibitem{Schulz:98}
H.~J. Schulz, G.~Cuniberti, and P.~Pieri, {\it {Fermi liquids and Luttinger
  liquids}},  \href{http://arxiv.org/abs/cond-mat/9807366}{{\tt
  cond-mat/9807366}}.

\bibitem{Maldacena:1997re}
J.~M. Maldacena, {\it {The Large N limit of superconformal field theories and
  supergravity}},  {\em Adv. Theor. Math. Phys.} {\bf 2} (1998) 231--252,
  [\href{http://arxiv.org/abs/hep-th/9711200}{{\tt hep-th/9711200}}].

\bibitem{Gubser:1998bc}
S.~S. Gubser, I.~R. Klebanov, and A.~M. Polyakov, {\it {Gauge theory
  correlators from noncritical string theory}},  {\em Phys. Lett. B} {\bf 428}
  (1998) 105--114, [\href{http://arxiv.org/abs/hep-th/9802109}{{\tt
  hep-th/9802109}}].

\bibitem{Witten:1998qj}
E.~Witten, {\it {Anti-de Sitter space and holography}},  {\em Adv. Theor. Math.
  Phys.} {\bf 2} (1998) 253--291,
  [\href{http://arxiv.org/abs/hep-th/9802150}{{\tt hep-th/9802150}}].

\bibitem{Aharony:1999ti}
O.~Aharony, S.~S. Gubser, J.~M. Maldacena, H.~Ooguri, and Y.~Oz, {\it {Large N
  field theories, string theory and gravity}},  {\em Phys. Rept.} {\bf 323}
  (2000) 183--386, [\href{http://arxiv.org/abs/hep-th/9905111}{{\tt
  hep-th/9905111}}].

\bibitem{Hartnoll:2008vx}
S.~A. Hartnoll, C.~P. Herzog, and G.~T. Horowitz, {\it {Building a Holographic
  Superconductor}},  {\em Phys. Rev. Lett.} {\bf 101} (2008) 031601,
  [\href{http://arxiv.org/abs/0803.3295}{{\tt arXiv:0803.3295}}].

\bibitem{Donos:2012js}
A.~Donos and S.~A. Hartnoll, {\it {Interaction-driven localization in
  holography}},  {\em Nature Phys.} {\bf 9} (2013) 649--655,
  [\href{http://arxiv.org/abs/1212.2998}{{\tt arXiv:1212.2998}}].

\bibitem{Ling:2014saa}
Y.~Ling, C.~Niu, J.~Wu, Z.~Xian, and H.-b. Zhang, {\it {Metal-insulator
  Transition by Holographic Charge Density Waves}},  {\em Phys. Rev. Lett.}
  {\bf 113} (2014) 091602, [\href{http://arxiv.org/abs/1404.0777}{{\tt
  arXiv:1404.0777}}].

\bibitem{An:2020tkn}
Y.-S. An, T.~Ji, and L.~Li, {\it {Magnetotransport and Complexity of
  Holographic Metal-Insulator Transitions}},  {\em JHEP} {\bf 10} (2020) 023,
  [\href{http://arxiv.org/abs/2007.13918}{{\tt arXiv:2007.13918}}].

\bibitem{Liu:2009dm}
H.~Liu, J.~McGreevy, and D.~Vegh, {\it {Non-Fermi liquids from holography}},
  {\em Phys. Rev. D} {\bf 83} (2011) 065029,
  [\href{http://arxiv.org/abs/0903.2477}{{\tt arXiv:0903.2477}}].

\bibitem{Ren:2010ha}
J.~Ren, {\it {One-dimensional holographic superconductor from AdS$_3$/CFT$_2$
  correspondence}},  {\em JHEP} {\bf 11} (2010) 055,
  [\href{http://arxiv.org/abs/1008.3904}{{\tt arXiv:1008.3904}}].

\bibitem{Liu:2011fy}
Y.~Liu, Q.~Pan, and B.~Wang, {\it {Holographic superconductor developed in BTZ
  black hole background with backreactions}},  {\em Phys. Lett. B} {\bf 702}
  (2011) 94--99, [\href{http://arxiv.org/abs/1106.4353}{{\tt
  arXiv:1106.4353}}].

\bibitem{Li:2012zzb}
R.~Li, {\it {Note on analytical studies of one-dimensional holographic
  superconductors}},  {\em Mod. Phys. Lett. A} {\bf 27} (2012) 1250001.

\bibitem{Alkac:2016otd}
G.~Alkac, S.~Chakrabortty, and P.~Chaturvedi, {\it {Holographic P-wave
  Superconductors in 1+1 Dimensions}},  {\em Phys. Rev. D} {\bf 96} (2017),
  no.~8 086001, [\href{http://arxiv.org/abs/1610.08757}{{\tt
  arXiv:1610.08757}}].

\bibitem{Maity:2009zz}
D.~Maity, S.~Sarkar, N.~Sircar, B.~Sathiapalan, and R.~Shankar, {\it
  {Properties of CFTs dual to Charged BTZ black-hole}},  {\em Nucl. Phys. B}
  {\bf 839} (2010) 526--551, [\href{http://arxiv.org/abs/0909.4051}{{\tt
  arXiv:0909.4051}}].

\bibitem{Faulkner:2012gt}
T.~Faulkner and N.~Iqbal, {\it {Friedel oscillations and horizon charge in 1D
  holographic liquids}},  {\em JHEP} {\bf 07} (2013) 060,
  [\href{http://arxiv.org/abs/1207.4208}{{\tt arXiv:1207.4208}}].

\bibitem{Hartnoll:2009sz}
S.~A. Hartnoll, {\it {Lectures on holographic methods for condensed matter
  physics}},  {\em Class. Quant. Grav.} {\bf 26} (2009) 224002,
  [\href{http://arxiv.org/abs/0903.3246}{{\tt arXiv:0903.3246}}].

\bibitem{Natsuume:2014sfa}
M.~Natsuume, {\it {AdS/CFT Duality User Guide}},
  \href{http://arxiv.org/abs/1409.3575}{{\tt arXiv:1409.3575}}.

\bibitem{Hartnoll:2016apf}
S.~A. Hartnoll, A.~Lucas, and S.~Sachdev, {\it {Holographic quantum matter}},
  \href{http://arxiv.org/abs/1612.07324}{{\tt arXiv:1612.07324}}.

\bibitem{Baggioli:2019rrs}
M.~Baggioli, {\it {Applied Holography}: {A Practical Mini-Course}},
  \href{http://arxiv.org/abs/1908.02667}{{\tt arXiv:1908.02667}}.

\bibitem{Baggioli:2021xuv}
M.~Baggioli, K.-Y. Kim, L.~Li, and W.-J. Li, {\it {Holographic Axion Model: a
  simple gravitational tool for quantum matter}},  {\em Sci. China Phys. Mech.
  Astron.} {\bf 64} (2021), no.~7 270001,
  [\href{http://arxiv.org/abs/2101.01892}{{\tt arXiv:2101.01892}}].

\bibitem{Hartnoll:2009ns}
S.~A. Hartnoll, J.~Polchinski, E.~Silverstein, and D.~Tong, {\it {Towards
  strange metallic holography}},  {\em JHEP} {\bf 04} (2010) 120,
  [\href{http://arxiv.org/abs/0912.1061}{{\tt arXiv:0912.1061}}].

\bibitem{Davison:2013txa}
R.~A. Davison, K.~Schalm, and J.~Zaanen, {\it {Holographic duality and the
  resistivity of strange metals}},  {\em Phys. Rev. B} {\bf 89} (2014), no.~24
  245116, [\href{http://arxiv.org/abs/1311.2451}{{\tt arXiv:1311.2451}}].

\bibitem{Anderson}
P.~W. Anderson, {\it {Hall effect in the two-dimensional Luttinger liquid}},
  {\em Phys. Rev. Lett.} {\bf 67} (1991), no.~15 2092.

\bibitem{coleman1996should}
P.~Coleman, A.~J. Schofield, and A.~M. Tsvelik, {\it {How should we interpret
  the two transport relaxation times in the cuprates?}},  {\em J. Phys: Cond.
  Matt.} {\bf 8} (1996), no.~48 9985--10015,
  [\href{http://arxiv.org/abs/cond-mat/9609009}{{\tt cond-mat/9609009}}].

\bibitem{Zhou:2015dha}
Z.~Zhou, J.-P. Wu, and Y.~Ling, {\it {DC and Hall conductivity in holographic
  massive Einstein-Maxwell-Dilaton gravity}},  {\em JHEP} {\bf 08} (2015) 067,
  [\href{http://arxiv.org/abs/1504.00535}{{\tt arXiv:1504.00535}}].

\bibitem{Mott:1968nwb}
N.~Nott, {\it {Metal-Insulator Transition}},  {\em Rev. Mod. Phys.} {\bf 40}
  (1968) 677--683.

\bibitem{Ling:2016dck}
Y.~Ling, P.~Liu, J.-P. Wu, and Z.~Zhou, {\it {Holographic Metal-Insulator
  Transition in Higher Derivative Gravity}},  {\em Phys. Lett. B} {\bf 766}
  (2017) 41--48, [\href{http://arxiv.org/abs/1606.07866}{{\tt
  arXiv:1606.07866}}].

\bibitem{Nishioka:2009zj}
T.~Nishioka, S.~Ryu, and T.~Takayanagi, {\it {Holographic
  Superconductor/Insulator Transition at Zero Temperature}},  {\em JHEP} {\bf
  03} (2010) 131, [\href{http://arxiv.org/abs/0911.0962}{{\tt
  arXiv:0911.0962}}].

\bibitem{Andrade:2013gsa}
T.~Andrade and B.~Withers, {\it {A simple holographic model of momentum
  relaxation}},  {\em JHEP} {\bf 05} (2014) 101,
  [\href{http://arxiv.org/abs/1311.5157}{{\tt arXiv:1311.5157}}].

\bibitem{Horowitz:2012gs}
G.~T. Horowitz, J.~E. Santos, and D.~Tong, {\it {Further Evidence for
  Lattice-Induced Scaling}},  {\em JHEP} {\bf 11} (2012) 102,
  [\href{http://arxiv.org/abs/1209.1098}{{\tt arXiv:1209.1098}}].

\bibitem{Horowitz:2012ky}
G.~T. Horowitz, J.~E. Santos, and D.~Tong, {\it {Optical Conductivity with
  Holographic Lattices}},  {\em JHEP} {\bf 07} (2012) 168,
  [\href{http://arxiv.org/abs/1204.0519}{{\tt arXiv:1204.0519}}].

\bibitem{Horowitz:2013jaa}
G.~T. Horowitz and J.~E. Santos, {\it {General Relativity and the Cuprates}},
  {\em JHEP} {\bf 06} (2013) 087, [\href{http://arxiv.org/abs/1302.6586}{{\tt
  arXiv:1302.6586}}].

\bibitem{Gouteraux:2016wxj}
B.~Gout\'eraux, E.~Kiritsis, and W.-J. Li, {\it {Effective holographic theories
  of momentum relaxation and violation of conductivity bound}},  {\em JHEP}
  {\bf 04} (2016) 122, [\href{http://arxiv.org/abs/1602.01067}{{\tt
  arXiv:1602.01067}}].

\bibitem{Baggioli:2016pia}
M.~Baggioli, B.~Gout\'eraux, E.~Kiritsis, and W.-J. Li, {\it {Higher derivative
  corrections to incoherent metallic transport in holography}},  {\em JHEP}
  {\bf 03} (2017) 170, [\href{http://arxiv.org/abs/1612.05500}{{\tt
  arXiv:1612.05500}}].

\bibitem{Li:2018vrz}
W.-J. Li and J.-P. Wu, {\it {A simple holographic model for spontaneous
  breaking of translational symmetry}},  {\em Eur. Phys. J. C} {\bf 79} (2019),
  no.~3 243, [\href{http://arxiv.org/abs/1808.03142}{{\tt arXiv:1808.03142}}].

\bibitem{Liu:2022bam}
Y.~Liu, X.-J. Wang, J.-P. Wu, and X.~Zhang, {\it {Alternating current
  conductivity and superconducting properties of the holographic effective
  theory}},  \href{http://arxiv.org/abs/2201.06065}{{\tt arXiv:2201.06065}}.

\bibitem{Zhong:2022mok}
Y.-Y. Zhong and W.-J. Li, {\it {Transverse Goldstone mode in holographic fluids
  with broken translations}},  \href{http://arxiv.org/abs/2202.05437}{{\tt
  arXiv:2202.05437}}.

\bibitem{Iqbal:2008by}
N.~Iqbal and H.~Liu, {\it {Universality of the hydrodynamic limit in AdS/CFT
  and the membrane paradigm}},  {\em Phys. Rev. D} {\bf 79} (2009) 025023,
  [\href{http://arxiv.org/abs/0809.3808}{{\tt arXiv:0809.3808}}].

\bibitem{Blake:2013bqa}
M.~Blake and D.~Tong, {\it {Universal Resistivity from Holographic Massive
  Gravity}},  {\em Phys. Rev. D} {\bf 88} (2013), no.~10 106004,
  [\href{http://arxiv.org/abs/1308.4970}{{\tt arXiv:1308.4970}}].

\bibitem{Donos:2014uba}
A.~Donos and J.~P. Gauntlett, {\it {Novel metals and insulators from
  holography}},  {\em JHEP} {\bf 06} (2014) 007,
  [\href{http://arxiv.org/abs/1401.5077}{{\tt arXiv:1401.5077}}].

\bibitem{Donos:2014cya}
A.~Donos and J.~P. Gauntlett, {\it {Thermoelectric DC conductivities from black
  hole horizons}},  {\em JHEP} {\bf 11} (2014) 081,
  [\href{http://arxiv.org/abs/1406.4742}{{\tt arXiv:1406.4742}}].

\bibitem{Baggioli:2017ojd}
M.~Baggioli and W.-J. Li, {\it {Diffusivities bounds and chaos in holographic
  Horndeski theories}},  {\em JHEP} {\bf 07} (2017) 055,
  [\href{http://arxiv.org/abs/1705.01766}{{\tt arXiv:1705.01766}}].

\bibitem{Baggioli:2016rdj}
M.~Baggioli, {\it {Gravity, holography and applications to condensed matter}},
  \href{http://arxiv.org/abs/1610.02681}{{\tt arXiv:1610.02681}}.

\bibitem{Donos:2013eha}
A.~Donos and J.~P. Gauntlett, {\it {Holographic Q-lattices}},  {\em JHEP} {\bf
  04} (2014) 040, [\href{http://arxiv.org/abs/1311.3292}{{\tt
  arXiv:1311.3292}}].

\bibitem{Ling:2015ghh}
Y.~Ling, {\it {Holographic lattices and metal\textendash{}insulator
  transition}},  {\em Int. J. Mod. Phys. A} {\bf 30} (2015), no.~28 1545013.

\bibitem{Ling:2015dma}
Y.~Ling, P.~Liu, C.~Niu, J.-P. Wu, and Z.-Y. Xian, {\it {Holographic
  Entanglement Entropy Close to Quantum Phase Transitions}},  {\em JHEP} {\bf
  04} (2016) 114, [\href{http://arxiv.org/abs/1502.03661}{{\tt
  arXiv:1502.03661}}].

\bibitem{Ling:2015epa}
Y.~Ling, P.~Liu, C.~Niu, and J.-P. Wu, {\it {Building a doped Mott system by
  holography}},  {\em Phys. Rev. D} {\bf 92} (2015), no.~8 086003,
  [\href{http://arxiv.org/abs/1507.02514}{{\tt arXiv:1507.02514}}].

\bibitem{Ling:2015exa}
Y.~Ling, P.~Liu, and J.-P. Wu, {\it {A novel insulator by holographic
  Q-lattices}},  {\em JHEP} {\bf 02} (2016) 075,
  [\href{http://arxiv.org/abs/1510.05456}{{\tt arXiv:1510.05456}}].

\bibitem{Ling:2016wyr}
Y.~Ling, P.~Liu, and J.-P. Wu, {\it {Characterization of Quantum Phase
  Transition using Holographic Entanglement Entropy}},  {\em Phys. Rev. D} {\bf
  93} (2016), no.~12 126004, [\href{http://arxiv.org/abs/1604.04857}{{\tt
  arXiv:1604.04857}}].

\bibitem{Baggioli:2014roa}
M.~Baggioli and O.~Pujolas, {\it {Electron-Phonon Interactions, Metal-Insulator
  Transitions, and Holographic Massive Gravity}},  {\em Phys. Rev. Lett.} {\bf
  114} (2015), no.~25 251602, [\href{http://arxiv.org/abs/1411.1003}{{\tt
  arXiv:1411.1003}}].

\bibitem{Baggioli:2016oqk}
M.~Baggioli and O.~Pujolas, {\it {On holographic disorder-driven
  metal-insulator transitions}},  {\em JHEP} {\bf 01} (2017) 040,
  [\href{http://arxiv.org/abs/1601.07897}{{\tt arXiv:1601.07897}}].

\bibitem{Baggioli:2016oju}
M.~Baggioli and O.~Pujolas, {\it {On Effective Holographic Mott Insulators}},
  {\em JHEP} {\bf 12} (2016) 107, [\href{http://arxiv.org/abs/1604.08915}{{\tt
  arXiv:1604.08915}}].

\bibitem{Donos:2014oha}
A.~Donos, B.~Gout\'eraux, and E.~Kiritsis, {\it {Holographic Metals and
  Insulators with Helical Symmetry}},  {\em JHEP} {\bf 09} (2014) 038,
  [\href{http://arxiv.org/abs/1406.6351}{{\tt arXiv:1406.6351}}].

\bibitem{Liu:2021stu}
P.~Liu and J.-P. Wu, {\it {Dynamic Properties of Two-Dimensional Latticed
  Holographic System}},  \href{http://arxiv.org/abs/2104.04189}{{\tt
  arXiv:2104.04189}}.

\bibitem{ziman2001electrons}
J.~M. Ziman, {\em Electrons and phonons: the theory of transport phenomena in
  solids}.
\newblock Oxford university press, 2001.

\bibitem{Mahajan:2013cja}
R.~Mahajan, M.~Barkeshli, and S.~A. Hartnoll, {\it {Non-Fermi liquids and the
  Wiedemann-Franz law}},  {\em Phys. Rev. B} {\bf 88} (2013) 125107,
  [\href{http://arxiv.org/abs/1304.4249}{{\tt arXiv:1304.4249}}].

\bibitem{Kuang:2017rpx}
X.-M. Kuang, E.~Papantonopoulos, J.-P. Wu, and Z.~Zhou, {\it {Lifshitz black
  branes and DC transport coefficients in massive Einstein-Maxwell-dilaton
  gravity}},  {\em Phys. Rev. D} {\bf 97} (2018), no.~6 066006,
  [\href{http://arxiv.org/abs/1709.02976}{{\tt arXiv:1709.02976}}].

\bibitem{Kim:2014bza}
K.-Y. Kim, K.~K. Kim, Y.~Seo, and S.-J. Sin, {\it {Coherent/incoherent metal
  transition in a holographic model}},  {\em JHEP} {\bf 12} (2014) 170,
  [\href{http://arxiv.org/abs/1409.8346}{{\tt arXiv:1409.8346}}].

\bibitem{Wu:2018zdc}
J.-P. Wu, X.-M. Kuang, and Z.~Zhou, {\it {Holographic transports from
  Born\textendash{}Infeld electrodynamics with momentum dissipation}},  {\em
  Eur. Phys. J. C} {\bf 78} (2018), no.~11 900,
  [\href{http://arxiv.org/abs/1805.07904}{{\tt arXiv:1805.07904}}].

\bibitem{Horowitz:1999jd}
G.~T. Horowitz and V.~E. Hubeny, {\it {Quasinormal modes of AdS black holes and
  the approach to thermal equilibrium}},  {\em Phys. Rev. D} {\bf 62} (2000)
  024027, [\href{http://arxiv.org/abs/hep-th/9909056}{{\tt hep-th/9909056}}].

\bibitem{Jansen:2017oag}
A.~Jansen, {\it {Overdamped modes in Schwarzschild-de Sitter and a Mathematica
  package for the numerical computation of quasinormal modes}},  {\em Eur.
  Phys. J. Plus} {\bf 132} (2017), no.~12 546,
  [\href{http://arxiv.org/abs/1709.09178}{{\tt arXiv:1709.09178}}].

\bibitem{Wu:2018vlj}
J.-P. Wu and P.~Liu, {\it {Quasi-normal modes of holographic system with Weyl
  correction and momentum dissipation}},  {\em Phys. Lett. B} {\bf 780} (2018)
  616--621, [\href{http://arxiv.org/abs/1804.10897}{{\tt arXiv:1804.10897}}].

\bibitem{Fu:2018yqx}
G.~Fu and J.-P. Wu, {\it {EM Duality and Quasinormal Modes from Higher
  Derivatives with Homogeneous Disorder}},  {\em Adv. High Energy Phys.} {\bf
  2019} (2019) 5472310, [\href{http://arxiv.org/abs/1812.11522}{{\tt
  arXiv:1812.11522}}].

\bibitem{Liu:2020lwc}
P.~Liu, C.~Niu, and C.-Y. Zhang, {\it {Linear instability of charged massless
  scalar perturbation in regularized 4D charged Einstein-Gauss-Bonnet anti
  de-Sitter black holes}},  {\em Chin. Phys. C} {\bf 45} (2021), no.~2 025111,
  [\href{http://arxiv.org/abs/2005.01507}{{\tt arXiv:2005.01507}}].

\end{thebibliography}\endgroup

\end{document}